\documentclass[12pt]{article}

\usepackage[english]{babel}
\usepackage{graphicx,rotating,epsfig}
\usepackage{a4p}

\def\GeV  {\ensuremath{\mathrm{ Ge\kern -0.1em V } }}
\def\GeVc2{\ensuremath{\mathrm{ Ge\kern -0.1em V }\kern -0.2em /c^2 }}
\newcommand{\MT}{\ensuremath{M_{\mathrm{top}}}}
\newcommand{\MTll}{\ensuremath{\MT^{\mathrm{di-l}}}}
\newcommand{\MTlj}{\ensuremath{\MT^{\mathrm{l+j}}}}
\newcommand{\MTjj}{\ensuremath{\MT^{\mathrm{all-j}}}}
\newcommand{\MW}{\ensuremath{M_{\mathrm{ W }}}}
\newcommand{\Pt}{\ensuremath{t}}
\newcommand{\Ptt}{\ensuremath{\Pt\bar\Pt}}
\newcommand{\RunI}{\hbox{Run-I}}
\newcommand{\RunII}{\hbox{Run-II}}

\begin{document}


\begin{center}
{\LARGE FERMI NATIONAL ACCELERATOR LABORATORY}
\end{center}

\begin{flushright}
       TEVEWWG/top 2005/02 \\
       FERMILAB-TM-2323-E \\  
       CDF Note 7727 V2 \\
       D\O\ Note 4892 \\
       hep-ex/0507091 \\
\end{flushright}

\vskip 1cm

\begin{center}
{\Huge \bf Combination of CDF and D\O\ Results \\[3mm]
                  on the Top-Quark Mass}
\vskip 1cm
{\Large
The CDF Collaboration, the D\O\ Collaboration, and \\[1mm]
the Tevatron Electroweak Working Group\footnote{
WWW access at {\tt http://tevewwg.fnal.gov}\\
The members of the TEVEWWG who contributed significantly to the
analysis described in this note are: \\
J.~F.~Arguin (arguin@fnal.gov)          
F.~Canelli (canelli@fnal.gov),          
R.~Demina (demina@fnal.gov),            
I.~Fleck (fleck@fnal.gov),              
D.~Glenzinski (douglasg@fnal.gov),      
E.~Halkiadakis (evah@fnal.gov),         
M.~W.~Gr\"unewald (mwg@fnal.gov),       
A.~Juste (juste@fnal.gov),              
T.~Maruyama (maruyama@fnal.gov),        
A.~Quadt (quadt@fnal.gov),              
E.~Thomson (thomsone@fnal.gov),         
C.~Tully (tully@fnal.gov),              
E.~W.~Varnes (varnes@fnal.gov),         
D.~O.~Whiteson (danielw@fnal.gov).      
}
}

\vskip 1cm

{\bf Abstract}

\end{center}

{
  
  The results on the measurements of the top-quark mass, based on the
  data collected by the Tevatron experiments CDF and D\O\ at Fermilab
  during {\RunI} from 1992 to 1996 and {\RunII} since 2001, are
  summarized. The combination of the published {\RunI} and preliminary
  {\RunII} results, taking correlated uncertainties properly into
  account, is presented.  The resulting preliminary world average for
  the mass of the top quark is: $\MT=172.7\pm 2.9~\GeVc2$, where the
  total error consists of a statistical part of $1.7~\GeVc2$ and a
  systematic part of $2.4~\GeVc2$.

  Compared to the combination prepared for the LP 2005 conference,
  this combination for the EPS-HEP 2005 conference includes additional
  published {\RunI} and preliminary {\RunII} measurements.

}

\vfill



\section{Introduction}

The experiments CDF and D\O, taking data at the Tevatron
proton-antiproton collider located at the Fermi National Accelerator
Laboratory, have made several direct experimental measurements of the
pole mass, $\MT$, of the top quark $\Pt$.  The published measurements
~\cite{Mtop1-CDF-di-l-PRLa, Mtop1-CDF-di-l-PRLb,
  Mtop1-CDF-di-l-PRLb-E, Mtop1-D0-di-l-PRL, Mtop1-D0-di-l-PRD,
  Mtop1-CDF-l+j-PRL, Mtop1-CDF-l+j-PRD, Mtop1-D0-l+j-old-PRL,
  Mtop1-D0-l+j-old-PRD, Mtop1-D0-l+j-new1, Mtop1-CDF-all-j-PRL,
  Mtop1-D0-all-j-PRL} are based on about $100\;\rm pb^{-1}$ of {\RunI}
data (1992-1996) while the results from the analyses of about
$320\;\rm pb^{-1}$ of {\RunII} date are
preliminary~\cite{Mtop2-CDF-di-l, Mtop2-CDF-l+j, Mtop2-CDF-l+j-new,
  Mtop2-CDF-ll-new, Mtop2-D0-di-l, Mtop2-D0-l+j, Mtop2-D0-l+j-new}.
They utilize all decay topologies\footnote{Decay channels with
  identification of tau leptons in the final state are presently under
  study for cross section and branching ratio measurements. They are
  not yet used for measurements of the top quark mass.} arising in
$\Ptt$ production given by the leptonic or hadronic decay of the W
boson occurring in top-quark decay: the di-lepton channel
(di-l)~\cite{Mtop1-CDF-di-l-PRLa, Mtop1-CDF-di-l-PRLb,
  Mtop1-CDF-di-l-PRLb-E, Mtop1-D0-di-l-PRL, Mtop1-D0-di-l-PRD,
  Mtop2-CDF-di-l, Mtop2-CDF-ll-new, Mtop2-D0-di-l}, the lepton+jets
channel (l+j)~\cite{Mtop1-CDF-l+j-PRL, Mtop1-CDF-l+j-PRD,
  Mtop1-D0-l+j-old-PRL, Mtop1-D0-l+j-old-PRD, Mtop1-D0-l+j-new1,
  Mtop2-CDF-l+j, Mtop2-CDF-l+j-new, Mtop2-D0-l+j, Mtop2-D0-l+j-new},
and the all-jets channel (all-j)~\cite{Mtop1-CDF-all-j-PRL,
  Mtop1-D0-all-j-PRL}. The lepton+jets channel yields the most precise
determination of $\MT$.  The recently presented preliminary
measurements in this channel by the CDF and D\O\ 
collaborations~\cite{Mtop2-CDF-l+j-new, Mtop2-D0-l+j-new} are based on
large {\RunII} data sets with well controlled systematic
uncertainties, each yielding a top quark mass precision similar to or
better than the previous {\RunI} world average~\cite{Mtop1-tevewwg04}.

This note reports on the combination of these measurements, using the
five published {\RunI} measurements~\cite{Mtop1-CDF-di-l-PRLb,
  Mtop1-CDF-di-l-PRLb-E, Mtop1-D0-di-l-PRD, Mtop1-CDF-l+j-PRD,
  Mtop1-D0-l+j-new1, Mtop1-CDF-all-j-PRL} combined
earlier~\cite{Mtop1-tevewwg04} and in particular including the most
recent preliminary {\RunII} measurements from
CDF~\cite{Mtop2-CDF-l+j-new,Mtop2-CDF-ll-new} and
D\O~\cite{Mtop2-D0-l+j-new}. The combination takes into account the
statistical and systematic uncertainties as well as the correlations
between systematic uncertainties, and replaces previous
combinations~\cite{Mtop1-tevewwg04,Mtop-tevewwg0506}.  The most
precise individual measurements of $\MT$ are now the preliminary
lepton+jets measurements from Run II.  These are
$173.5^{+4.1}_{-4.0}\;\rm GeV/c^2$ (CDF, \cite{Mtop2-CDF-l+j-new}) and
$169.5 \pm 4.7\;\rm GeV/c^2$ (D\O, \cite{Mtop2-D0-l+j-new}). These
have weights in the new $\MT$ combination of 36\% and 33\%,
respectively.

\section{Measurements}

The eight measurements of $\MT$ to be combined are listed in
Table~\ref{tab:inputs}. The preliminary {\RunII} D\O\ measurement in
the lepton+jets channel constrains the jet energy scale from an
in-situ calibration, based on the hadronic $W\rightarrow q q^\prime$
invariant mass in the $\Ptt$ events.  The preliminary {\RunII} CDF
measurement in the lepton+jets channel constrains the jet energy scale
simultaneously from external studies (calorimeter-track comparisons on
$E/p$ from single isolated tracks) as well from an in-situ calibration
using the W-boson mass.

For the combination procedure the preliminary {\RunII} CDF lepton+jets
channel is split into two separate measurements with identical central
value and fully correlated statistical and systematic errors. Only the
jet energy scale uncertainty is uncorrelated.  One measurement,
(l+j)$_e$, is associated with an energy scale uncertainty of
$3.1~\GeV$ from the external calibration, which is fully correlated
with the energy scale uncertainty of other results.  The other
measurement, (l+j)$_i$, is associated with an energy scale uncertainty
of $4.2~\GeV$, estimated to be the contribution from the in-situ
calibration, which is uncorrelated with any CDF or D\O\ result.  The
combination of these two measurements yields identical central value,
statistical error and systematic error, and a total jet energy scale
uncertainty of $2.5~\GeV$, as quoted in~\cite{Mtop2-CDF-l+j-new} for
the preliminary {\RunII} CDF measurement in the lepton+jets channel.

Besides central values and statistical uncertainties, the systematic
errors arising from various sources are reported in
Table~\ref{tab:inputs}.  For each measurement, the individual error
contributions are combined in quadrature.  In order of decreasing
importance, the systematic error sources are:
\begin{itemize}
\item Jet energy scale (JES): The systematics for jet energy scale
  include the uncertainties on the absolute jet energy corrections,
  calorimeter stability, underlying event and relative jet energy
  corrections. Since the jet energy scale uncertainty is the largest
  uncertainty in all channels, dominating the overall precision of
  this combination, the various components of this uncertainty have
  been studied quantitatively and are grouped into contributions,
  correlated or uncorrelated between the channels, the two
  experiments, and the data-taking periods ({\RunI}, {\RunII}). Based
  on studies of the correlations of the JES subcomponents in the
  {\RunII} data, also the systematic uncertainties of the {\RunI}
  measurements are, retrospectively, split into various components so
  that the correlations of those subcomponents can be better taken
  into account. The respective error groups are as follows:
  
  {\bf iJES:} The component of the jet energy scale originating from
  in-situ calibration procedures, here using the $W \rightarrow
  qq^\prime$ invariant mass in the preliminary {\RunII} l+j channels
  from CDF and D\O, is labeled iJES. For D\O\ Run-II this component
  includes both the statistical error from the in-situ calibration and
  the systematic error from a possible additional $p_T$ dependence of
  the data to Monte Carlo ratio of the jet energy scale. The
  correlation due to using the same method based on the same
  assumptions and tested with the same MC is presently estimated to be
  negligible.
  
  {\bf aJES and bJES:} The components of the jet energy scale covering
  aspects of the $b$-jet energy scale.  The part labeled bJES
  includes fragmentation, color flow and semi-leptonic b decay
  fractions, and is treated as fully correlated between all channels
  of all experiments.  The additional part labeled aJES for D\O\ 
  {\RunII} contains the uncertainty arising from the detector
  electromagnetic/hadronic calibration of b-jets versus light-quark
  jets which was negligible in {\RunI} due to a different calorimeter
  read-out. The uncertainties assigned to the {\RunI} measurements are
  based on studies of the {\RunII} data, and subtracted quadratically
  from the total JES error quoted in the {\RunI} publications.
  
  {\bf cJES:} The correlated part of the remaining, external jet
  energy scale uncertainty (cJES) from the external calibration
  includes uncertainties from fragmentation and out-of-cone showering
  corrections and is correlated between all channels of all
  experiments.
  
  {\bf dJES:} The part of the jet energy scale uncertainty correlated
  between measurements within the same data-taking period (either
  {\RunI} or {\RunII}) but not between experiments, arising from the
  calibration data samples.

  {\bf rJES:} The remaining external jet energy scale uncertainty
  (rJES) summarizes uncertainties mainly from the calorimeter
  response, relative response of different calorimeter sections,
  multiple interactions for CDF and contributions from the underlying
  event. It is treated as correlated between all channels of a given
  experiment independent of data-taking period, but not between
  experiments.
  
\item Model for signal (signal): The systematics for the signal model
  include initial and final state radiation effects, b-tagging bias,
  dependence upon parton distribution functions as well as variations
  in $\Lambda_{\mathrm{QCD}}$.

\item Model for background (BG): The background model includes the
  effect of varying the background fraction attributed to QCD
  multi-jet production with fake leptons and missing $E_T$.  It also
  includes the estimates of the effects of varying the fragmentation
  scale and fragmentation model. In {\RunI}, the scale was varied from
  $Q^2=\MW^2$ to $Q^2=\langle p_t\rangle^2$ in VECBOS~\cite{VECBOS}
  simulations of W+jets production, and ISAJET~\cite{ISAJET}
  fragmentation was used instead of HERWIG~\cite{HERWIG5}
  fragmentation. In D\O\ {\RunII}, ALPGEN~\cite{ALPGEN} and
  PYTHIA~\cite{PYTHIA4,PYTHIA5,PYTHIA6} are compared, and the scale is
  varied from the default $Q^2 = \MW^2 + \sum p_{T,j}^2$ to $Q^2 =
  \langle p_{T,j}^2\rangle$.

\item Uranium noise and multiple interactions (UN/MI): This
  uncertainty includes uncertainties arising from uranium noise in the
  D\O\ calorimeter and from multiple interactions overlapping signal
  events. CDF includes the systematic uncertainty due to multiple
  interactions in the uncorrelated part of the JES contribution from
  external calibration. For D\O\ in {\RunII}, the shorter integration
  time, plus the fact that the in-situ JES calibration includes these
  contributions, results in this uncertainty becoming negligibly small
  and therefore not considered for this preliminary result.
  
\item Method for mass fitting (fit): This systematic uncertainty takes
  into account the finite sizes of Monte Carlo samples used for
  fitting, impact of jet permutations, and other fitting biases.  In
  the CDF {\RunI} lepton+jets analysis, the systematic uncertainty due
  to finite Monte Carlo statistics is included in the statistical
  uncertainty.
  
\item Monte Carlo generator (MC): The systematic uncertainty on the
  Monte Carlo generator provides an estimate of the sensitivity to the
  simulated physics model by comparing HERWIG to PYTHIA or to ISAJET.
  In the D\O\ analyses, the systematic uncertainty associated with the
  comparison of HERWIG to ISAJET ({\RunI}) or replacing 30\% of the
  ALPGEN $\Ptt$ simulation by an ALPGEN $\Ptt+jets$ simulation
  ({\RunII}) is included in the signal model uncertainty.

\end{itemize}
Further studies on the systematic errors, in particular on the
breakdown of the various contributions to the jet energy scale
uncertainties and the use of leading order MC, and the correlations
are necessary to achieve better understanding and will be pursued in
the future.  The described procedure with the quoted numbers represent
our current, preliminary understanding of the various error sources
and their correlations.

\begin{table}[t]
\begin{center}
\renewcommand{\arraystretch}{1.30}
\begin{tabular}{|l||rrr|rr||rrr|r|}
\hline       
       & \multicolumn{5}{|c||}{{\RunI} published} & \multicolumn{4}{|c|}{{\RunII} preliminary} \\
\hline
       & \multicolumn{3}{|c|}{ CDF } & \multicolumn{2}{|c||}{ D\O\ }
       & \multicolumn{3}{|c|}{ CDF } & \multicolumn{1}{|c|}{ D\O\ } \\
\hline       
       & all-j &   l+j &  di-l &   l+j &  di-l & (l+j)$_i$ & (l+j)$_e$ & di-l & l+j \\
\hline                         
\hline                         
Result & 186.0 & 176.1 & 167.4 & 180.1 & 168.4 & 173.5 & 173.5 & 165.3 & 169.5 \\
\hline                         
\hline                         
Stat.  &  10.0 &   5.1 &  10.3 &   3.6 &  12.3 &   2.7 &   2.7 &   6.3 &   3.0 \\
\hline                         
\hline                         
iJES   &   0.0 &   0.0 &   0.0 &   0.0 &   0.0 &   4.2 &   0.0 &   0.0 &   3.3 \\
aJES   &   0.0 &   0.0 &   0.0 &   0.0 &   0.0 &   0.0 &   0.0 &   0.0 &   0.9 \\
bJES   &   0.6 &   0.6 &   0.8 &   0.7 &   0.7 &   0.6 &   0.6 &   0.8 &   0.7 \\
cJES   &   3.0 &   2.7 &   2.6 &   2.0 &   2.0 &   0.0 &   2.0 &   2.2 &   0.0 \\
dJES   &   0.3 &   0.7 &   0.6 &   0.0 &   0.0 &   0.0 &   0.0 &   0.0 &   0.0 \\
rJES   &   4.0 &   3.4 &   2.7 &   2.5 &   1.1 &   0.0 &   2.3 &   1.4 &   0.0 \\
                               
Signal &   1.8 &   2.6 &   2.8 &   1.1 &   1.8 &   1.1 &   1.1 &   1.5 &   0.3 \\
MC     &   0.8 &   0.1 &   0.6 &   0.0 &   0.0 &   0.2 &   0.2 &   0.8 &   0.0 \\
UN/MI  &   0.0 &   0.0 &   0.0 &   1.3 &   1.3 &   0.0 &   0.0 &   0.0 &   0.0 \\
BG     &   1.7 &   1.3 &   0.3 &   1.0 &   1.1 &   1.2 &   1.2 &   1.6 &   0.7 \\
Fit    &   0.6 &   0.0 &   0.7 &   0.6 &   1.1 &   0.6 &   0.6 &   0.6 &   0.6 \\
\hline                         
Syst.  &   5.7 &   5.3 &   4.9 &   3.9 &   3.6 &   4.6 &   3.5 &   3.6 &   3.6 \\
\hline                         
\hline                         
Total  &  11.5 &   7.3 &  11.4 &   5.3 &  12.8 &   5.3 &   4.4 &   7.3 &   4.7 \\
\hline
\end{tabular}
\end{center}
\caption[Input measurements]{Summary of the eight measurements of
$\MT$ performed by CDF and D\O. All numbers are in $\GeVc2$. Note that
the preliminary {\RunII} CDF measurement in the lepton+jets channel is
split into two measurements in order to treat the correlations of the
jet energy scale uncertainties properly. For each measurement, the
corresponding column lists experiment and channel, central value and
contributions to the total error, namely statistical error and
systematic errors arising from various sources defined in the text.
Overall systematic errors and total errors are obtained by combining
individual errors in quadrature.}
\label{tab:inputs}
\end{table}

\clearpage

\section{Combination}

In the combination, the error contributions arising from different
sources are uncorrelated between measurements.  The correlations of
error contributions arising from the same source are as follows:
\begin{itemize} 
\item uncorrelated: statistical error, fit error, iJES error;
\item 100\% correlated within each experiment: rJES error, UN/MI error;
\item 100\% correlated within each experiment for the same data-taking
period (either {\RunI} or {\RunII}): aJES, dJES;
\item 100\% correlated within each channel: BG error;
\item 100\% correlated between all measurements: bJES error, cJES
error, signal error, MC error.
\end{itemize}
Note that the jet energy scale uncertainty from the in-situ
calibration (iJES) in the preliminary {\RunII} lepton+jets measurements
from CDF and D\O\ are treated as uncorrelated with any other
measurement.  All uncertainties except iJES are treated fully
correlated between the two preliminary {\RunII} CDF lepton+jets
measurements.  The resulting matrix of global correlation coefficients
is listed in Table~\ref{tab:correl}.

\begin{table}[t]
\begin{center}
\renewcommand{\arraystretch}{1.30}
\begin{tabular}{|ll||rrr|rr||rrr|r|}
\hline       
   &   & \multicolumn{5}{|c||}{{\RunI} published} & \multicolumn{4}{|c|}{{\RunII} preliminary} \\
\hline
   &   & \multicolumn{3}{|c|}{ CDF } & \multicolumn{2}{|c||}{ D\O\ }
       & \multicolumn{3}{|c|}{ CDF } & \multicolumn{1}{|c|}{ D\O\ } \\
\hline       
   &  &   all-j &   l+j &  di-l &   l+j &  di-l & (l+j)$_i$ & (l+j)$_e$ & di-l & l+j \\
\hline
\hline
CDF-I & all-j   &   1.00&       &       &       &       &       &       &      &     \\
CDF-I & l+j     &   0.32&   1.00&       &       &       &       &       &      &     \\
CDF-I & di-l    &   0.19&   0.29&   1.00&       &       &       &       &      &     \\
\hline
D\O-I & l+j     &   0.14&   0.26&   0.15&   1.00&       &       &       &      &     \\
D\O-I & di-l    &   0.07&   0.11&   0.08&   0.16&   1.00&       &       &      &     \\
\hline
\hline
CDF-II & (l+j)$_i$& 0.04&   0.12&   0.06&   0.10&   0.03&   1.00&       &      &     \\
CDF-II & (l+j)$_e$& 0.35&   0.54&   0.29&   0.29&   0.11&   0.45&   1.00&      &     \\
CDF-II & di-l   &   0.19&   0.28&   0.18&   0.17&   0.10&   0.06&   0.30&  1.00&     \\
\hline
D\O-II & l+j    &   0.02&   0.07&   0.03&   0.07&   0.02&   0.07&   0.08&  0.03&  1.00\\
\hline
\end{tabular}
\end{center}
\caption[Global correlations between input measurements]{Matrix of
global correlation coefficients between the measurements of $\MT$.}
\label{tab:correl}
\end{table}

The measurements are combined using a program implementing a numerical
$\chi^2$ minimization as well as the analytic BLUE
method~\cite{Lyons:1988, Valassi:2003}. The two methods used are
mathematically equivalent, and are also equivalent to the method used
in an older combination~\cite{TM-2084}, and give identical results for
the combination. In addition, the BLUE method yields the decomposition
of the error on the average in terms of the error categories specified
for the input measurements~\cite{Valassi:2003}.

\section{Results}

The combined value for the top-quark mass is:
\begin{eqnarray}
\MT & = & 172.7 \pm 2.9~\GeVc2\,,
\end{eqnarray}
where the total error of $2.9~\GeVc2$ contains the following
components: a statistical error of $1.7~\GeVc2$; and systematic error
contributions of: total JES $2.0~\GeVc2$, signal $0.9~\GeVc2$,
background $0.9~\GeVc2$, UN/MI $0.3~\GeVc2$, fit $0.3~\GeVc2$, and MC
$0.2~\GeVc2$, for a total systematic error of $2.4~\GeVc2$.

\begin{figure}[p]
\begin{center}
\includegraphics[width=0.8\textwidth]{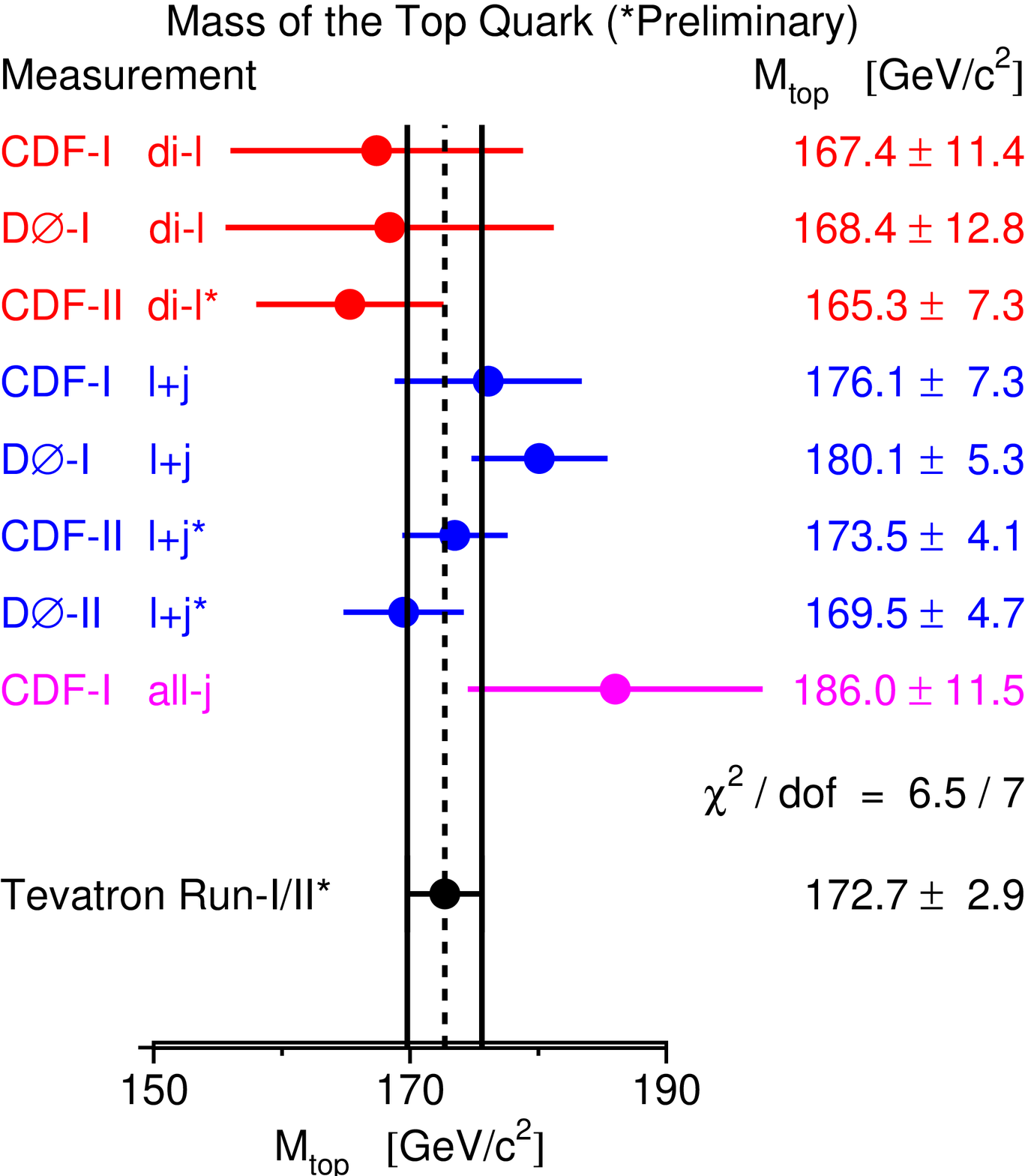}
\end{center}
\caption[Comparison of the measurements of the top-quark mass]
{Comparison of the measurements of the top-quark mass and their
average.}
\label{fig:mtop-bar-chart} 
\end{figure}

The $\chi^2$ of this average is 6.45 for 7 degrees of freedom,
corresponding to a probability of 49\%, showing that all measurements
are in good agreement with each other which can also be seen in
Figure~\ref{fig:mtop-bar-chart}.  The pull of each measurement with
respect to the average and the weight of each measurement in the
average are reported in Table~\ref{tab:stat}. Note that the weight of
the CDF-I lepton+jets measurement is negative. In general, this
situation can occur if the correlation of two measurements of a
physical quantity is larger than the ratio of their errors. In case
the correlation coefficient $r$ becomes equal to the error ratio, the
weight of the less precise measurement becomes zero as it does not
improve the combination. While a weight=0 means that the lower
accuracy measurement is ignored, a negative weight implies that this
particular result does contribute to lowering the variance of the
final answer. In our combination, the CDF-I lepton+jets measurements
and the externally calibrated CDF-II lepton+jets measurement turn out
to be 54\% correlated (see Table~\ref{tab:correl}), while the Run-II
measurement has almost half the error of the Run-I measurement (see
Table~\ref{tab:inputs}).  See Reference~\cite{Lyons:1988} for further
discussion of negative weights.

\begin{table}[t]
\begin{center}
\renewcommand{\arraystretch}{1.30}
\begin{tabular}{|l||rrr|rr||rrr|r|}
\hline       
       & \multicolumn{5}{|c||}{{\RunI} published} & \multicolumn{4}{|c|}{{\RunII} preliminary} \\
\hline
       & \multicolumn{3}{|c|}{ CDF } & \multicolumn{2}{|c||}{ D\O\ }
       & \multicolumn{3}{|c|}{ CDF } & \multicolumn{1}{|c|}{ D\O\ } \\
\hline       
       &  all-j  &  l+j    &  di-l   &   l+j   &  di-l    &(l+j)$_i$&(l+j)$_e$&   di-l & l+j \\
\hline
\hline
Pull   & $+1.19$ & $+0.51$ & $-0.48$ & $+1.67$ &  $-0.34$ & $+0.18$ & $+0.24$ & $-1.11$ & $-0.86$ \\
\hline
Weight [\%]
       & $+ 1.0$ & $- 0.2$ & $+ 1.1$ & $+18.8$ &  $+ 2.1$ & $+18.6$ & $+17.4$ & $+ 8.0$ & $+33.3$ \\
\hline
\end{tabular}
\end{center}
\caption[Pull and weight of each measurement]{Pull and weight of each
  measurement in the average. See Reference~\cite{Lyons:1988} for a
  discussion of negative weights.}
\label{tab:stat} 
\end{table} 

In addition, a combination of the eight measurements in three physical
observables, the top quark mass in the di-lepton channel, $\MTll$, the
lepton+jets channel, $\MTlj$, and the all-jets channel, $\MTjj$, has
been performed. The results of this combination, obtained with a
$\chi^2$ of 2.64 for 5 degrees of freedom corresponding to a
probability of 76\%, are shown in Table~\ref{tab:three_observables}.
Please note that those results differ from a naive combination, where
only the one measurement in the all-jets channel is considered for
$\MTjj$, the five measurements in the lepton+jets channel are
considered for $\MTlj$ and the three measurements in the dilepton
channel are combined into $\MTll$. In our combination the correlations
between systematic uncertainties of all measurements, also the ones
made in different channels, are properly taken into account in a
global fit and can change the central fit values and errors of the
result, yielding smaller errors on the three combined results than
three independent combinations.

\begin{table}[h]
\begin{center}
\renewcommand{\arraystretch}{1.30}
\begin{tabular}{|l||c|rrr|}
\hline
Parameter & Value & \multicolumn{3}{|c|}{Correlations} \\
\hline
\hline
$\MTjj~ [\GeVc2]$ & $185.0\pm          10.9$ & 1.00 &      &      \\
$\MTlj~~[\GeVc2]$ & $173.5\pm\phantom{0}3.0$ & 0.22 & 1.00 &      \\
$\MTll~ [\GeVc2]$ & $165.0\pm\phantom{0}5.8$ & 0.15 & 0.31 & 1.00 \\
\hline
\end{tabular}
\end{center}
\caption[Input measurements]{Summary of the combination of the eight
measurements by CDF and D\O\ in terms of three physical quantities,
the top quark mass in the di-lepton, lepton+jets and all-jets
channel. }
\label{tab:three_observables}
\end{table}

\section{Summary}

A preliminary combination of measurements of the mass of the top
quark, $\MT$, from the Tevatron experiments CDF and D\O\ is presented.
The combination includes five published {\RunI} measurements and three
preliminary {\RunII} measurements.  Taking into account statistical and
systematic errors including their correlations, the preliminary
Tevatron and thus the world-average result is: $\MT= 172.7 \pm
2.9~\GeVc2$.  The mass of the top quark is now known with an accuracy
of 1.7\%.

\clearpage

\bibliographystyle{tevewwg}
\bibliography{run2mtop0507}

\end{document}